\documentclass{epl}

\usepackage{amsmath}
\usepackage{amssymb}
\usepackage[dvips]{color}

\newcommand{\abs}[1]{\ensuremath{\left|#1\right|}}
\newcommand{\dd}{\mathrm{d}}
\title{Nonstationary dephasing of two level systems}
 \author{J. Schriefl\inst{1,2} \and M. Clusel\inst{1}
    \and D. Carpentier\inst{1} \and P. Degiovanni\inst{1}}
 \institute{
   \inst{1} CNRS-Laboratoire de Physique de l'{\'E}cole normale
            sup{\'e}rieure de Lyon,
            46, All{\'e}e d'Italie, 69007 Lyon, France\\
   \inst{2} Institut f{\"u}r Theoretische
            Festk{\"o}rperphysik, Universit{\"a}t Karlsruhe,
            76128 Karlsruhe, Germany
           }
 \shortauthor{J. Schriefl \etal}


\begin{document}
\maketitle
\begin{abstract}
We investigate the influence of  nonstationary $1/f^\mu$ noise,
produced by interacting defects, on
a quantum two-level system.
Adopting a simple phenomenological model for this noise
we describe exactly the corresponding dephasing in various regimes.
The nonstationarity and
pronounced non-Gaussian features of this noise induce new anomalous
dephasing scenarii.
Beyond a history-dependent critical coupling strength
the dephasing time exhibits a strong dependence on the age of the noise
and the decay of coherence is not exponential.
\end{abstract}

  The question of the phase coherence of a quantum two level system
(TLS) in a noisy environment has motivated numerous recent works. In
particular it is of  crucial interest in the context of quantum
computing with  solid state quantum bits\cite{Makhlin2001}.
In this case it is now believed that
dephasing of the TLS by a low-frequency ($1/f$) noise is a limiting
mechanism which certainly deserves further studies.
 The microscopic origin of these low-frequency fluctuations
depends on the considered system and is mostly not
understood\cite{Nakamura2002}.
Hence, in previous works on this problem
the $1/f$ environment was
modeled  phenomenologically  either by a  set of harmonic modes
({\it e.g.} spin-Boson model)\cite{Makhlin2003} or by
an ensemble of independent bistable
fluctuators (Dutta-Horn model)\cite{weissman88,Paladino2002}.
 The first model assumes a Gaussian distribution for the fluctuations
and stationarity of the noise. As this low frequency noise appears
to be an essentially out of equilibrium phenomenon, this last
hypothesis is questionable\cite{Galperin2003}. The Dutta-Horn
model does not suffer from this restriction\cite{Paladino2002}, but
assumes the presence of a broad distribution of relaxation times
for the independent fluctuators, or equivalently a flat distribution
for their random energy barriers. However, both the nature of these
fluctuators and the origin of these long relaxation times are beyond
the current experimental knowledge.
We consider the
interactions between the fluctuators as a possible source for these
slow dynamics.
In this letter, we explore the consequences of
this interaction on the nonstationarity of the dephasing by
considering a simple phenomenological model for a strongly coupled
cluster of fluctuators.

 Let us thus consider a dissipative quantum two-level system (qubit)
 described by the Hamiltonian :
\begin{equation}\label{eq:Hamiltonian}
{\cal H} = \epsilon\: \sigma^{z} + \Delta\:
\sigma^{x} - \frac{1}{2} X \: \sigma^{z}  \; ,
\end{equation}
where $\epsilon$ and $\Delta$ are control parameters which we
assume to be time-independent and $X$ describes the fluctuating
extra bias (noise) induced by the environment.
 In this letter, we focus on the case $\Delta =0$. However,
extensions to the case $\Delta \neq 0$ and a low frequency noise $X$
may be treated along the lines of ref. \cite{Makhlin2003}. 
Within a  Dutta-Horn
model\cite{Paladino2002,Galperin2003},
the noise $X$
consists in the sum of contributions from independent
fluctuators, modeled as random two-level systems coupled to an
equilibrium bath ({\it e.g.} the phonons). This bosonic bath
induces an elastic or electromagnetic coupling 
between the qubit and each 
 fluctuator 
but also mutual couplings {\it of same origin} between the fluctuators
which are usually neglected.
 This interaction 
depends on the nature of the coupling to the
bath : both  elastic strain (phonons) and dipolar electric coupling
correspond to a long-range dipolar interaction\cite{Black77,footnote99}
 The motivation to consider such 
couplings  between the fluctuators comes in part from the
analogous physics of TLS in
amorphous solids
: the relevance of these TLS couplings was stressed by
Yu and Leggett \cite{Yu88}, and found recent
experimental evidence in the dielectric response of amorphous
films\cite{Osheroff2003}. These interactions are a possible source of
broad distribution of 
 relaxation times of the bath : in particular,
 using the language of pseudospins for the fluctuators, a strong
ferromagnetic coupling between an 
ensemble of $N$ spins forces these spins to tunnel simultaneously,
therefore reducing exponentially with the size $N$  
the tunneling amplitude of the
cluster. Such a nonlinear relation between the size of a cluster and
the corresponding relaxation time would naturally lead to a broad
distribution for these times.
Moreover, for randomly
located fluctuators elastically coupled, we expect the interactions
 to be random and frustrating \cite{Black77}. It is then
natural to consider the noise produced by a mesoscopic ensemble of $N$
interacting fluctuators\cite{footnote1}.
We expect these interactions and the associated
frustration to bring very slow relaxational dynamics for the cluster.
 However this problem is a
formidable task, still unsolved in the case of classical spin
glasses. We therefore
adopt a phenomenological approach and focus on
the expected general consequences of its physics.

\begin{figure}[htbp]
\begin{center}
\input{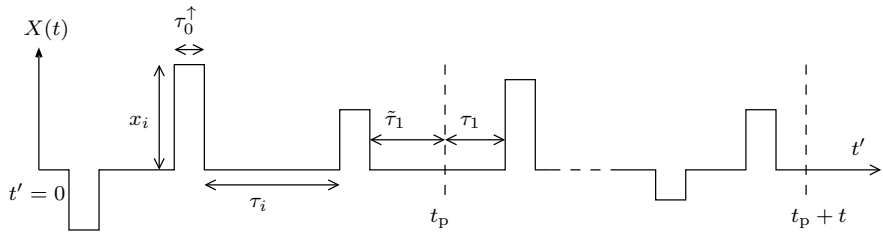}
\caption{Representation of a noise configuration $X(t)$}
\label{fig:telegraphe}
\end{center}
\end{figure}

As in the case of independent fluctuators \cite{Galperin2003,Paladino2002}
we will focus on a classical description of the noise $X(t)$,
considering an incoherent
environment and neglecting back-action effects of the qubit.
A common picture to describe the evolution of a glassy
system is a random walk in phase space with broadly distributed time
intervals $\tau$ between the 'flips of the environment'. Hence,
by analogy with the phase space trap models of classical glassy
systems\cite{Bouchaud1992}, we consider a  general random intermittent
noise, with randomly distributed
heights $x_{i}$ and durations $\tau_{i}^{\uparrow}$ of the plateaus,
and waiting
times $\tau_{i}$ between them (see Fig.\ref{fig:telegraphe}).
For simplicity, we will focus on the
case where the $\tau_{i}^{\uparrow}$ are much shorter than the $\tau_{i}$,
and given by their average $\tau^{\uparrow}_{0}$. On time scales $\tau\gg \tau_0^\uparrow$,
the plateau is seen as a spike shifting the phase by $x_i\tau_0^\uparrow$.
Moreover, to focus on dephasing
and eliminate the drift of the accumulated phase of the qubit,
the distribution $\mathcal{P}(x)$ is assumed to have zero mean and
a finite width $g^{2} =\left< x^{2} \right>(\tau^{\uparrow}_{0})^{2}$: It
will be taken as Gaussian without loss of generality.
Finally, we consider an algebraic
distribution of waiting times  $\tau_{i}$ characterized by an exponent
$\mu$:
\begin{equation}
\label{eq:defP}
P(\tau)=\frac{\mu}{\tau_{0}}
\left(\frac{\tau_{0}}{\tau_{0}+\tau} \right)^{1+\mu}
\quad ,\quad
\mathcal{P}(x)=\frac{\tau_{0}^{\uparrow}}{g\sqrt{2\pi}}
~\exp\left(-\frac{(x\tau^{\uparrow}_{0})^{2}}{2g^{2}} \right)
\end{equation}
where $\tau_{0}$ is a microscopic time. The slow algebraic decay
of $P(\tau)$ thereby implies divergences of all moments
$\overline{\tau^n} = \int_0^\infty d\tau P(\tau) \tau^n$ with $n \geq \mu$.
It turns out that three different classes of $\mu$ have to be distinguished:
(i) If both $\overline{\tau}$ and $\overline{\tau^2}$ are finite $(\mu >2)$,
the dephasing induced by the noise in our model and
within the usual model of Poissonian telegraph
noise\cite{Galperin2003,Paladino2002} are identical.
Thus, we will call it the {\it Poissonian class}.
However, in the case of slow dynamics, \textit{i.e.} for $\mu<2$,
the dephasing scenario differs considerably from the Poissonian class.
(ii) For $1<\mu<2$, when $\overline{\tau}$ is still finite, but already
the second moment $\overline{\tau^2}$ diverges, the waiting times $\tau_i$
start  to fluctuate strongly around their average $\overline{\tau}$
and important corrections with respect to dephasing due to Poissonian
class arise. (iii) For $0<\mu<1$,
the divergence of the first moment $\overline{\tau}$ makes it
even impossible to define a characteristic time between two consecutive
spikes. In this case the two-point noise correlation
function decays as
\begin{equation}
  \label{eq:Correlation}
\overline{X(t+t') X(t)} -\overline{X(t)} ~\overline{X(t+t')}
\simeq \overline{X(t)}^2
\left(t / t'\right)^{1-\mu} \; ,
\end{equation}
where the overline denotes an average over
realizations of $X(t)$. Eq.(\ref{eq:Correlation}) implies
that the power spectrum exhibits a $1/f^\mu$ divergence at low
frequencies. Note that the explicit dependence
of the correlator \eqref{eq:Correlation} on $t$ implies a
nonstationarity of the noise and consequently a time dependence of
the amplitude of the noise \cite{footnote3}. In our model, a single
source of noise with $\mu=1$ can lead to a $1/f$ power spectrum
whereas infinitely
many independent fluctuators are necessary in the Dutta-Horn
approach \cite{Paladino2002}.
Within this latter approach, nonstationary of the noise comes from the choice of
initial conditions for the slow fluctuators.
In our model, nonstationarity finds its origin in the
very slow dynamics underlying the noise, not in a
specific choice of an initial condition.

The nonstationarity of the noise
leads us to pay special attention to the dependence of dephasing
on the age of the noise (denoted by $t_{p}$ below), {\it i.e.} a history
dependence.
To be definite we consider situations where the noise has been initialized
at time $t'=0$.
The coupling to the qubit is turned on after a time $t_{\mathrm{p}}$, the preparation time
of the qubit.
Consequently, the two qubit states $|\pm\rangle$ accumulate
a random relative phase between
time $t_{\mathrm{p}}$ and $t_{\mathrm{p}}+t$:
$\Phi(t_{\mathrm{p}},t)= \int_{t_{\mathrm{p}}}^{t_{\mathrm{p}}+t}\dd t'~X(t')$.
The corresponding dephasing factor is defined by the
average over many realizations of noise
\begin{equation}
D(t_{\mathrm{p}},t)=\overline{\exp \left(i\ \Phi(t_{\mathrm{p}},t) \right)}
\end{equation}
and  the dephasing time $\tau_{\phi}$  as its characteristic
decay time: $D(t_{\mathrm{p}},\tau_{\phi}) = e^{-1}$.
In the following we will always consider preparation times $t_{\mathrm{p}}$ that are
large compared to the microscopic time $\tau_0$.
Under this assumption dephasing properties derived below do not
depend on the
imposed initial condition of the noise (at $t'=0$). In particular,
the noise is stationary for $\mu>2$, as in the case of
usual Poissonian noise
treated in \cite{Paladino2002,Galperin2003}.

Within the model (\ref{eq:defP}), the accumulated phase
$\Phi(t_{p},t)$ performs a {\it continuous time random walk}
(CTRW)\cite{haus87}. The explicit dependence of $D(t_{p},t)$ on
$t_{\mathrm{p}}$ can be accounted for using renewal
theory\cite{Feller}.
The times  $\tau_{i}$ between two successive spikes being independent
from each other (and thus of
the history of the noise),  the $t_{p}$-dependence of $\Phi (t_{p},t)$ and
$D(t_{p},t)$ are consequences of the $t_{p}$-dependence
of the distribution of $\tau_{1}$, the time interval between
 $t_{p}$ and the first
subsequent spike (see Fig.~\ref{fig:telegraphe}).
Indeed, the previous spike did not occur at $t_{p}$ but
at some time $t_{p}-\tilde{\tau}_{1}$.
Hence $\tilde\tau_{1}+\tau_1$ is
distributed according to $P(\tau)$ in ($\ref{eq:defP}$), whereas
$\tau_{1}$ is distributed according to
a new distribution $H_{t_{p}}(\tau_{1})$
which explicitly  depends on
$t_{p}$. As we show, this distribution $H_{t_{p}}(\tau_{1})$ contains all
information about the history of the noise.
Following general ideas from renewal theory we
separate noise configurations that have their
first spike at $t_{\mathrm{p}}+\tau_1$ from the others to obtain an expression
for $H_{t_{\mathrm{p}}}(\tau_{1})$ :
\begin{equation}
\label{eq:defHtp}
H_{t_{\mathrm{p}}}(\tau_{1}) =
P(t_{\mathrm{p}}+\tau_{1})+
\int_{0}^{t_{\mathrm{p}}}P(\tau_{1}+\tilde{\tau}_{1})
S(t_{\mathrm{p}}-\tilde{\tau}_{1})\,\dd\tilde{\tau}_{1},
\end{equation}
where $S(t)$ denotes the renewal distribution, \textit{i.e.} the density distribution
of spike at time $t$. The behavior of $S(t)$ follows from another renewal equation,
\begin{equation}
\label{eq:defS}
S(t)=P(t)+\int_{0}^{t}P(\tau)S(t-\tau)\,\dd \tau  ,
\end{equation}
which states that a flip occurring at time $t$ is either the
first one, or that a previous flip occurred at time $t-\tau$, where
$\tau$ is distributed according to (\ref{eq:defP}). Denoting by $L[f]$
the Laplace transform of the function $f$, Eq.(\ref{eq:defS}) can
be rewritten as $L[S]=L[P]/(1-L[P])$.
For $\mu>2$, $S$ is constant and
$H_{t_{p}}(\tau_{1})$ coincides with $P(\tau_{1})$ at times $t_p\gg \tau_0$.
After a transcient regime, the noise has lost memory of its initial condition:
the probability to flip to another state does not depend on the history of the noise.
For $\mu<2$, the time  dependence of $S(t)$
implies an explicit history dependence of
$H_{t_{\mathrm{p}}}(\tau_{1})$. For instance, for $1 < \mu < 2$,
$S(t)\simeq \overline{\tau}^{-1}(1+\left(\tau_{0} / t )^{\mu-1} \right)$, and
the average value of $\langle\tau_1\rangle$
increases as $t_{\mathrm{p}}^{2-\mu}$.
In the case $\mu<1$ of diverging mean value, $\overline{\tau} = \infty$,
the density of spikes $S(t)$ decreases as $1/t^{1-\mu}$ and
$H_{t_{\mathrm{p}}}$ differs therefore considerably from $P$.

\begin{figure}
\begin{minipage}[t]{0.5\linewidth}
\centering
\includegraphics[width=6cm]{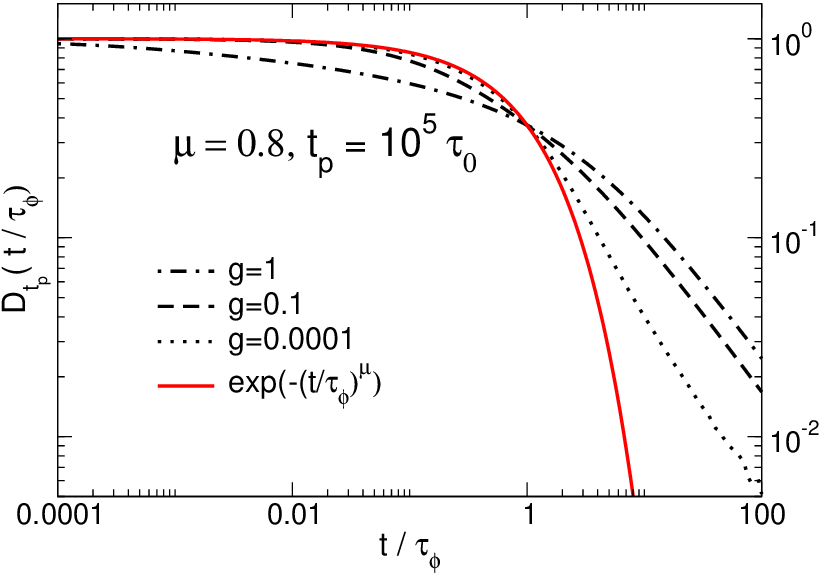}\vspace*{0.8cm}\\
\includegraphics[width=6cm]{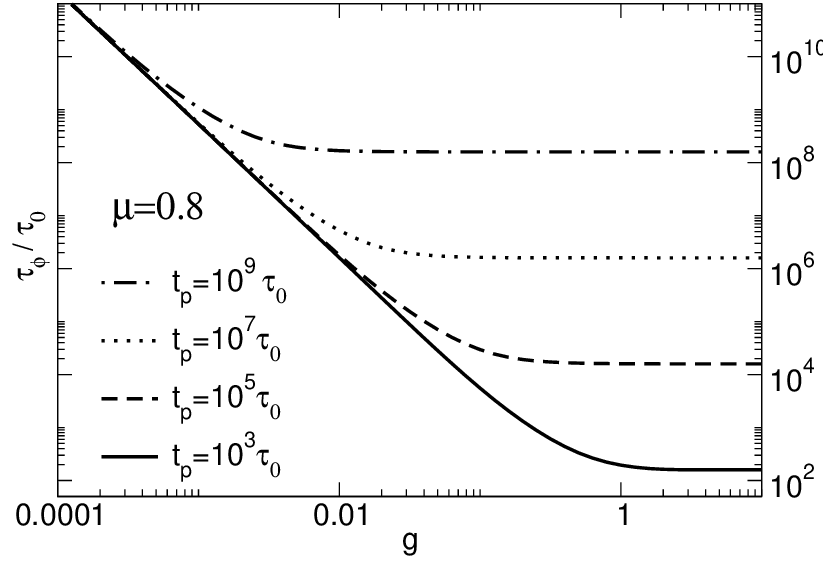}
\end{minipage}
\hfill
\begin{minipage}[t]{0.5\linewidth}
\centering
\includegraphics[width=6cm]{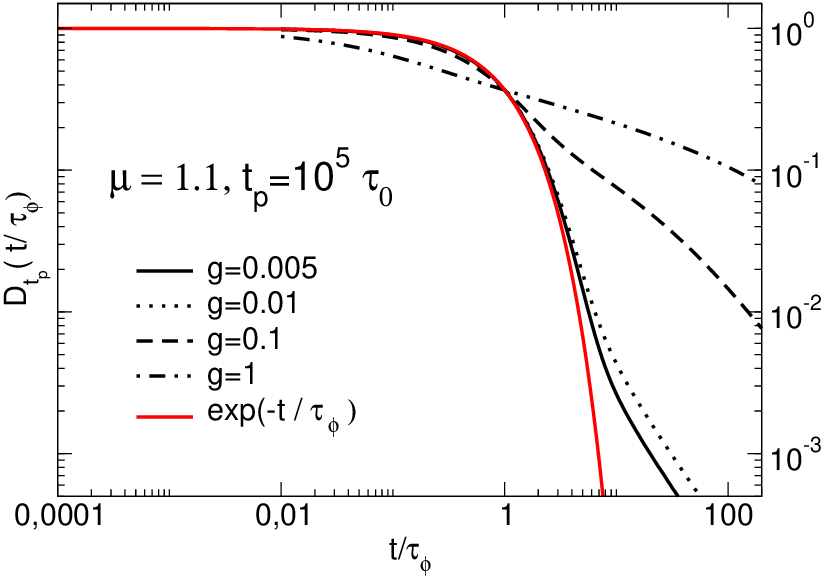}\vspace*{0.8cm}\\
\includegraphics[width=6cm]{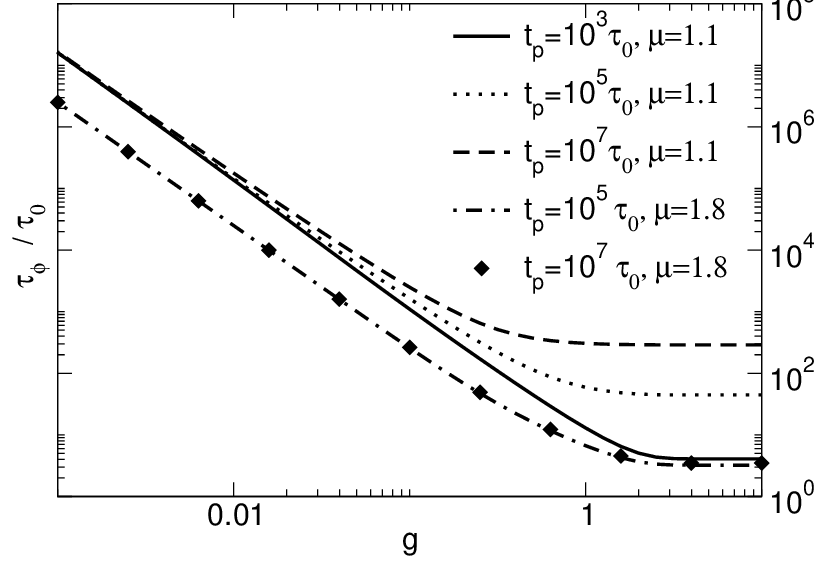}
\end{minipage}
\caption{Dephasing factor and $\tau_\phi$ obtained by numerical
inversion of (\ref{eq:Ptp2}) for $\mu < 1$ (left) and $1<\mu<2$ (right).
For weak coupling, $g<g_c(t_{\mathrm{p}})$, the decay is exponential for $t<\tau_\phi$.
For times $t>\tau_\phi$ $(\mu<1)$ - $t\gg \tau_\phi$ $(1<\mu<2)$ -
and for strong coupling the decay is algebraic.
For strong coupling and $\mu<1$ the dephasing time exhibits a
$t_{\mathrm{p}}$-dependence, which disappears as $\mu$ increases to higher values.
Note the explicit $t_{\mathrm{p}}$-dependence of the critical coupling $g_c$ in the
case $\mu<1$.
}
\label{fig:res}
\end{figure}

Now understanding  the origin of aging, we decompose  noise configurations
into those without any event between $t_{\mathrm{p}}$
and $t_{\mathrm{p}}+t$, and those with at least one event in the
same interval, the first of which occurring at time $t_{\mathrm{p}}+\tau$. In the
latter case, the first event contributes a factor
$\left< e^{ix\tau_{0}^{\uparrow}} \right>$,
reinitializing simultaneously the noise
at $t_{\mathrm{p}}+\tau$:
\begin{equation}
\label{eq:Dyson}
D(t_{\mathrm{p}},t)
=\Pi_0(t_{\mathrm{p}},t)+f(g)\,\int_0^t H_{t_{\mathrm{p}}}(\tau)
D(0,t-\tau)\,\dd \tau  .
\end{equation}
In this equation,
$f(g)=\left< e^{ix\tau_{0}^{\uparrow}} \right> =\exp (-g^{2}/2)$ is
the characteristic function of $\mathcal{P}(x)$ and
$\Pi_0(t_{\mathrm{p}},t)=\int_t^{+\infty}\dd\tau\,H_{t_{\mathrm{p}}}(\tau)$
denotes the probability that no spike occurs between $t_{\mathrm{p}}$ and
$t_{\mathrm{p}}+t$. More conveniently,
denoting by $L[D_{t_{\mathrm{p}}}](s)$ the Laplace transform  of
$D(t_{\mathrm{p}},t)$ considered as a function of $t$,
and specializing (\ref{eq:Dyson}) to $t_{\mathrm{p}}=0$, we find
$L[D_{0}](s)=s^{-1} (1-L[P](s))/(1-f(g)L[P](s))$.
Plugging it back into
the Laplace transform of (\ref{eq:Dyson}) we obtain
\begin{equation}
\label{eq:Ptp2}
L[D_{t_{\mathrm{p}}}](s)
=\frac{1}{s}\left(1-\frac{(1-f(g))L[H_{t_{\mathrm{p}}}]}{1-f(g)L[P]} \right).
\end{equation}
This expression is one of the
central results of this letter.
Indeed, we can apply a numerical Laplace Transform inversion to
obtain the complete behavior of the dephasing factor
$D(t_{\mathrm{p}},t)$.  Some of these results are shown in Fig.~\ref{fig:res}.
 The behavior of $D(t_{\mathrm{p}},t)$ is shown
for $\mu=0.8$, $\mu=1.1$ and both weak and strong coupling.

We now turn to a discussion of the results we can derive exactly from Eq.(\ref{eq:Ptp2}).
As discussed above, the dephasing scenario differs
qualitatively and quantitatively depending
on the value of $\mu$.
Generally, two regimes appear as a function of the coupling constant
$g$, separated by a critical coupling constant $g_c$:
While the dephasing time depends on $g$ in the weak coupling regime,
$g<g_c$, it saturates for $g>g_c$ as a function of $g$
(see Fig.~\ref{fig:res}).
As expected, it turns out that the
critical coupling strength $g_c$ is of order 1 for the Poissonian
class $(\mu>2)$ as well as for $1<\mu<2$.
However, in the case $\mu\leq 1$, we find that the range
of the strong coupling regime increases with the age of the noise, \textit{i.e.}
$g_c$ decreases as a function of $t_{\mathrm{p}}$:
$g_c(t_{\mathrm{p}}) = \lambda(\mu) (\tau_0/t_{\mathrm{p}})^{\mu/2}$.
This implies that any qubit surrounded by a
noise with $0<\mu <1$ will eventually end up in the strong coupling
regime.

In the following we will discuss in detail the decay of $D(t_{\mathrm{p}},t)$
as a function of $t$ and the scaling laws of $\tau_\phi$ as functions of $g$
for the three different classes of $\mu$ and
for both weak and strong coupling.
In particular, we will compare our results with those for the
Poissonian class. For $\mu >2$, $H_{t_{\mathrm{p}}}(\tau)\simeq P(\tau)$ and
from (\ref{eq:Ptp2})
$D(t_{\mathrm{p}},t) \simeq D(0,t)$. As expected, after
a transcient regime ($t_p\lesssim \tau_0$),  $D(t_{\mathrm{p}},t)$ becomes
independant of $t_{\mathrm{p}}$. The results of
\cite{Paladino2002,Galperin2003} for a single fluctuator
are then recovered for the Poissonian class.

For $1<\mu<2$, strong corrections with respect to
the Poissonian class $(\mu>2)$ occur in the strong coupling
regime, $g>1$.
In that regime, the decay law of $D(t_{\mathrm{p}},t)$ is no longer
an exponential but becomes algebraic
$D(t_{\mathrm{p}},t) \simeq
[\tau_{0}/(\tau_{0}+t)]^{\mu-1} -
[\tau_{0}/(\tau_{0}+t+t_{\mathrm{p}})]^{\mu-1}$.
The dephasing time thereby exhibits an explicit
$t_{\mathrm{p}}$-dependence,
$\tau_{\phi} \simeq \tau_0 [1/e + (\tau_0/t_{\mathrm{p}})^{\mu-1}]^{-1/(\mu-1)}$.
Note that this $t_{\mathrm{p}}$ dependence is only important for values of
$\mu$ close
to one and disappears as $\mu$ approaches higher values
(see Fig.~\ref{fig:res}).
For weak coupling, $g<1$, the initial decay of $D(t_{\mathrm{p}},t)$ is
very well described within a Gaussian approximation (second cumulant expansion).
The decay is exponential, $D(t_{\mathrm{p}},t) = \exp(-t/\tau_\phi)$, and
$\tau_{\phi}$ scales as $\tau_{\phi} \simeq \tau_{0} g^{-2}$.
Corrections with respect to Poissonian class are only subdominant for
$t\lesssim\tau_\phi$
and disappear as $\mu$ increases to higher values.
However, for times large compared to the dephasing time,
$t \gtrsim \tau_\phi \log(\tau_\phi/\tau_0)$, the above Gaussian
approximation breaks down and the decay crosses over to a much slower power law
(see Fig.~\ref{fig:res}).

In the case of $0<\mu <1$,
the decay of coherence differs considerably from the Poissonian class
in all
regimes, due to the absence of a characteristic time scale in the
waiting times distribution.
For weak coupling $g<g_{c}(t_{\mathrm{p}})$
the decay for $t\lesssim\tau_\phi$ is
accurately described by an exponential,
$D(t_{\mathrm{p}},t) \simeq\exp[-(t/\tau_\phi)^\mu]$ with
$\tau_{\phi} = \tau_{0} A(\mu) g^{-2/\mu }$. Dephasing is therefore
insensitive to the preparation time $t_{\mathrm{p}}$. This
result can be recovered using the previously mentionned
Gaussian approximation. However, we emphasize the anomalous
scaling of $\tau_\phi$ as a function of $g$ (see Fig.~\ref{fig:res}).
But $D(t_{\mathrm{p}},t)$ is not fully described within this approximation.
First of all, even at weak coupling, the Gaussian approximation breaks down for times
$t>\tau_{\phi}$. In  this regime, the exponential decay is replaced
by a much slower algebraic one:
$D(t_{\mathrm{p}},t) \simeq B(\mu) ~ (t/\tau_{\phi})^{-\mu}$.
This new behavior can be understood by noting
that in this case and for $t > \tau_{\phi}$
$D(t_{\mathrm{p}},t)\simeq
D(t_{\mathrm{p}},\tau_{\phi})~\Pi_{0}(\tau_{\phi },t-\tau_{\phi})
\simeq (t/\tau_{\phi}-1)^{-\mu}$ .
The second term of (\ref{eq:Dyson})
corresponds to the anomalous random walk spreading of
the phase $\Phi$: it
leads to the exponential decay (see above) which is subleading
beyond $\tau_{\phi}$. Hence, the leading term corresponds to the
contribution of the noise configurations that did not change
between $\tau_{\phi}$ and $t$. This situation of a main contribution
induced by rare configurations is analogous to the physics
of Griffiths singularities in disordered systems.
In the strong coupling regime $g>g_{c}(t_{\mathrm{p}})$, $D(t_{\mathrm{p}},t)$
is dominated by the first term of Eq.(\ref{eq:Dyson}). In this limit,
the distribution of the phase of the qubit starts spreading over
$[0,2\pi]$. Hence, most noise configurations produce a
vanishing contribution to $D(t_{\mathrm{p}},t)$
and the whole average is dominated by those noise configurations that do not
evolve during the experiment.
The physics of this strong coupling regime is closely related to the
physics of mean-field trap models of glassy
materials\cite{Bouchaud1992}, since the difference between quenched and
annealed disorder is irrelevant in this case.
The dephasing time then saturates as a function of $g$ and becomes
proportional to $t_{\mathrm{p}}$ as seen in Fig.~\ref{fig:res}.
Note that, at strong coupling, the Gaussian approximation breaks even down for short
times and the decay of $D(t_{\mathrm{p}},t)$ is algebraic:
it decays as $D(t_{\mathrm{p}},t) \simeq 1 - \sin(\pi\mu)/(\pi\mu)
(t/t_{\mathrm{p}})^{1-\mu}$.
For longer times, $t>t_{\mathrm{p}}$, it crosses over to a much slower decay,
$D(t_{\mathrm{p}},t) \simeq (t_{\mathrm{p}}/t)^{\mu}$.

Finally, in the marginal case $\mu=1$ of pure $1/f$ noise
the above analysis is confirmed qualitatively. We mainly find
logarithmic corrections to the above results, e.g. the critical coupling
scales as
$g_c(t_{\mathrm{p}}) = \left[2(\tau_0/t_{\mathrm{p}})
\ln(t_{\mathrm{p}}/\tau_0)\right]^{1/2}$.
For weak coupling, $g<g_c(t_{\mathrm{p}})$, the dephasing time is
insensitive to $t_{\mathrm{p}}$
and scales as $\tau_\phi \simeq 2\tau_0\abs{\ln g}/g^2$, whereas for strong
coupling it depends algebraically on $t_{\mathrm{p}}$.


In summary, we have studied decoherence of a qubit due to the noise generated by
a slow collective environment.
Within a simple phenomenological model, we have derived exact expressions
for the dephasing factor in various regimes. The crucial consequences
of the nonstationarity of the noise are first the appearance of a
history dependent coupling $g_{c}(t_{\mathrm{p}})$ which separate the weak and
strong coupling regimes and, second, a non exponential decrease of the
dephasing factor $D(t_{\mathrm{p}},t)$. Depending on the broadness
($\mu$) of the distribution of relaxation times of the environment, this
nonexponential decay appears either before ($\mu <1$) or after
$\tau_{\phi}$, but is a clear signature of the nonstationarity of
the noise ({\it i.e.} $\mu <2$ in our model). In a more refined
description of a collective noise source consisting of a collection of
coupled clusters, we expect the nonstationarity of the dephasing to
be related to the number of contributing clusters. Similarly to the
studies of Paladino {\it et al.} \cite{Paladino2002} for usual
telegraphic fluctuators, the nonstationary dephasing in our model will survive provided
the phase $\Phi(t_{\mathrm{p}},t)$ is dominated by a few strongly coupled slow
clusters. In the opposite limit of many contributing clusters, the
usual Gaussian stationary dephasing should be recovered.
Thus the search for such nonstationarity in the dephasing of simple solid
state qubits would be of main interest for the correct characterization
of the source of $1/f$ noise in these samples. Note however that
dephasing is experimentally studied by repeated interference
experiments without noise reinitialization
in between. The dephasing is then characterized by a time-average
 instead of the average over configurations
$D(t_{\mathrm{p}},t)$.
In the case of a nonstationary noise these two expressions will
differ (non ergodicity for $0<\mu\leq 1$)
and special care should be given in analyzing the experimental
results.

J. Schriefl thanks Yu. Makhlin for very useful discussions.
P. Degiovanni thanks  the Institute for Quantum Computing (Waterloo) and
Boston University for support and hospitality during completion of
this work.


\begin{thebibliography}{999}
\bibitem{footnote1}
Note that the physics associated with huge degeneracy of the
classical ground state should be irrelevant in such
mesoscopic clusters.
\end{thebibliography}
\end{document}